\documentclass[a4paper,11pt]{article}
\pdfoutput=1 

\usepackage{jheppub} 

\usepackage[T1]{fontenc} 

\usepackage{slashed}
\usepackage[utf8]{inputenc}

\newcommand{\bwt}{\begin{widetext}}
\newcommand{\ewt}{\end{widetext}}

\def\ord{\mathcal{O}}
\def\mO{\mathcal{O}}
\def\TeV{~{\mbox{TeV}}}

\def\GeV{~{\mbox{GeV}}}
\def\MeV{~{\mbox{MeV}}}

\def\mL{\mathcal{L}}

\title{\boldmath Constraining CP-violating electron-gluonic operators}

\author[a,b,c]{Kingman Cheung,}
\author[d]{Wai-Yee Keung,}
\author[a]{Ying-nan Mao,}
\author[a]{Chen Zhang}

\affiliation[a]{Physics Division, National Center for Theoretical Sciences, Hsinchu, Taiwan 300}
\affiliation[b]{Department of Physics, National Tsing Hua University, Hsinchu 300, Taiwan}
\affiliation[c]{Division of Quantum Phases and Devices, School of Physics,
Konkuk University, Seoul 143-701, Republic of Korea}
\affiliation[d]{Department of Physics, University of Illinois at Chicago, Illinois 60607 USA}

\emailAdd{cheung@phys.nthu.edu.tw}
\emailAdd{keung@uic.edu}
\emailAdd{ynmao@cts.nthu.edu.tw}
\emailAdd{czhang@cts.nthu.edu.tw}

\preprint{NCTS-PH/1903}

\abstract{We present an analysis of constraints on two types of CP-odd electron-gluonic operators
$(\bar{e}i\gamma^5 e)G_{\mu\nu}^a G^{a\mu\nu}$ and $(\bar{e}e)G_{\mu\nu}^a \tilde{G}^{a\mu\nu}$ from current
and future electric dipole moment (EDM) experiments. The recent result from the ACME experiment using ThO molecules is used to
derive an impressive lower bound on the effective scale for $(\bar{e}i\gamma^5 e)G_{\mu\nu}^a G^{a\mu\nu}$
at $8\TeV$, assuming a QCD one-loop factor and no helicity suppression from new physics. One interesting aspect is
that $(\bar{e}i\gamma^5 e)G_{\mu\nu}^a G^{a\mu\nu}$ contributes to the observable EDM in ThO experiment mainly through
CP-odd electron-nucleon interaction rather than direct electron EDM which arises from three-loop running and matching.
For $(\bar{e}e)G_{\mu\nu}^a \tilde{G}^{a\mu\nu}$ the current bound is much weaker and suffers from large uncertainties.
We also discuss the QCD running and matching for the CP-odd electron-gluonic operators and give an estimate of
the relevant nucleon matrix elements and uncertainties that are needed in the calculation.

}

\begin{document}
\maketitle
\flushbottom

\section{Introduction}
\label{sec:intro}

Tests of fundamental discrete symmetries have proven to be crucial for the establishment
of the Standard Model (SM) of particle physics. The discovery of parity violation~\cite{Lee:1956qn,Wu:1957my}
in weak interactions entails the introduction of a chiral gauge theory, in which the
generation of elementary fermion masses is tied to the electroweak symmetry breaking
mechanism in a nontrivial manner~\cite{Weinberg:1967tq}. The time-reversal symmetry, or equivalently CP invariance
(assuming an exact CPT symmetry), turns out to be even more mysterious. Currently all
experimentally observed CP violation can be well accommodated by a single Cabibbo–Kobayashi–Maskawa (CKM) phase, while
the effective CP-violating $\theta$ angle in the strong interaction is constrained to be
vanishingly small for unknown reasons~\cite{Tanabashi:2018oca}. The CP-violation from the CKM phase is however too
small to explain the observed matter-antimatter asymmetry of the universe. Given that these
various clues do not point to a clear picture undoubtedly, it is reasonable to be open-minded
about the search for new sources of CP-violation at all frontiers and the interpretation
of the results thereof.

Recently the ACME Collaboration has set a new constraint on the electron electric dipole
moment (EDM) using ThO molecules~\cite{Andreev:2018ayy}:
\begin{align}
|d_e|<1.1\times 10^{-29}e\cdot\text{cm}\,\,(90\% \text{C.L.})
\label{eq:acme}
\end{align}
leading to stringent bounds on new sources of CP-violation. For example, this result can be
translated into a stringent bound on the imaginary part of the electron Yukawa coupling~\cite{Altmannshofer:2015qra,Egana-Ugrinovic:2018fpy} (assuming
no accidental cancellation): $|\text{Im}\kappa_e|<2\times 10^{-3}$ where $\kappa_e$ denotes the ratio of
the complex electron Yukawa to its SM value. This is remarkable since at high energy colliders
the current direct search for Higgs decaying to electrons can only constrain the real part of the electron
Yukawa to be less than a few hundred times its SM value~\cite{Altmannshofer:2015qra,Egana-Ugrinovic:2018fpy}. More generally, if beyond the SM (BSM) CP
violation occurs at some high scale $\Lambda$, its effect can be parametrized by CP-violating higher-dimensional
operators in the SM effective field theory (SMEFT), which in turn can be bounded by electron EDM
measurements. Due to selection rules, at dimension-six level only a few operators contribute to the renormalization
of the electron dipole operators at one-loop~\cite{Panico:2018hal}. To maximally exploit the stringent
constraint in Eq.~\eqref{eq:acme} two-loop contribution from dimension-six operators and one-loop
contribution from dimension-eight operators may also be considered, which are expected to
deliver comparable constraints on the EFT scale $\Lambda$~\cite{Panico:2018hal}.

In this work we aim to constrain CP-violating interactions between electron and gluon, which
can be parametrized in SMEFT using dimension-eight operators. There are a number of reasons
why we are interested in these lepton-gluonic operators. Phenomenologically they can lead to
clean signatures at hadron colliders or in lepton-flavor-violation
measurements~\cite{Potter:2012yv,Hayreter:2013vna,Cai:2018cog} (if lepton-flavor is not
conserved). More interestingly, as we will show, one of the CP-violating electron-gluonic
operators is bounded by electron EDM measurements using ThO
molecules mainly due to its contribution to CP-odd electron-nucleon interactions rather than
direct contribution to electron EDM. In fact, what the ACME experiment really constrains
is the following combination of direct electron EDM and contribution from CP-odd electron nucleon
coupling~\cite{Cesarotti:2018huy,Chupp:2014gka,Dzuba:2011}
\begin{align}
d_{\text{exp}}=d_e+kC_S,\quad\quad k\approx 1.6\times 10^{-15}\GeV^2 e\cdot\text{cm}
\end{align}
Here $C_S$ is the coefficient of the CP-odd electron-nucleon operator $(\bar{e}i\gamma^5 e)(\bar{N}N)$.
In the SMEFT, up to dimension-six, $C_S$ receives contribution from certain CP-odd four-fermion operators
which involve two electrons and two quarks, due to quark contents in the nucleons. Such four-fermion operators
may arise in extended Higgs sector or leptoquark models~\cite{Barr:1991yx,Barr:1992cm,Fuyuto:2017xup,Fuyuto:2018scm,Dekens:2018bci}.
Since there also exist gluon contents in the
nucleons it is natural to ask whether constraints can be put on CP-odd electron-gluonic operators. In the
literature such operators have been considered in the following contexts. First, if there exist CP-odd $(\bar{e}i\gamma^5 e)\bar{Q}Q$
operators where $Q$ denotes a heavy quark ($c,b,t$) in the SM, then when we integrate out the heavy quark
in the EFT, CP-odd electron-gluonic operators could be generated via matching~\cite{Shifman:1978zn}. Second, in supersymmetric models
the CP-odd electron-gluonic operator $(\bar{e}i\gamma^5 e)G_{\mu\nu}^a G^{a\mu\nu}$ can be generated through
quark and squark loops~\cite{Pilaftsis:2002fe,Ellis:2008zy,Yanase:2018qqq}, which however suffers from helicity suppression. In this work, however, our interest
will be constraining in a model-independent manner, CP-odd electron-gluonic operators that are of independent new physics origin (i.e. not
generated by matching from CP-odd $eeQQ(Q=c,b,t)$ operators) and are potentially not suppressed by electron
Yukawa.

Moreover, we will also investigate whether meaningful bounds can be put on the $(\bar{e}e)G_{\mu\nu}^a \tilde{G}^{a\mu\nu}$
type operator using current and future EDM measurements. At electron-nucleon interaction level, this operator is reduced to
$\left(\bar{e}e\right)\left(\bar{N}\textrm{i}\gamma^5N\right)$ so that it is proportional to the averaged nucleon spin $\langle s_z\rangle$.
However, for ThO molecules used in ACME experiments, in both Th- and O-nuclei, the protons and neutrons are all paired which leads
to $\langle s_z\rangle=0$ for both kinds of nuclei \footnote{The ACME experiment chose the isotope $^{232}\textrm{Th}^{16}\textrm{O}$ which have the largest natural abundance ($\sim1$), and thus the pollution from other isotopes are ignored.}.
That means through only a tree level analysis, the ACME experiments cannot
be used to constrain the $(\bar{e}e)G_{\mu\nu}^a \tilde{G}^{a\mu\nu}$ type operator. We must turn to higher-order analysis and other
materials (for example, some heavy atoms), as shown in the text.

The paper is organized as follows. In Section~\ref{sec:uvr} the relevant electron-gluonic operators are listed
in the SMEFT context, and their UV realizations are discussed. Due to renormalization group running
electron-gluonic operators could mix into four-fermion operators, which is treated in Section~\ref{sec:orm},
including possible threshold matching effects. In Section~\ref{sec:oee} we discuss observable EDM effects
from CP-odd electron-gluonic operators, including both direct contribution to electron EDM and contribution
to CP-odd electron-nucleon interactions. The latter turns out to be the dominant effect. Then in Section~\ref{sec:ccfp}
based on the formulas obtained we present the bounds on the coefficient of CP-odd electron-gluonic
operators using current measurements and also make projections for future experiments. Section~\ref{sec:dnc}
gives the discussion and conclusion.

\section{CPV electron-gluonic operators and UV realizations}
\label{sec:uvr}
In the SMEFT, at the dimension-eight level, CP-odd electron-gluonic interactions can be introduced through the Lagrangian
\begin{align}
\mL_{eg}^{CPV}=C_g\mO_g^{GI}+\tilde{C}_g\tilde{\mO}_g^{GI}
\label{eq:OGI}
\end{align}
Here $C_g$ and $\tilde{C}_g$ are dimensionless real numbers, and
the $SU(3)_c\times SU(2)_L\times U(1)_Y$-invariant operators $\mO_g^{GI}$
and $\tilde{\mO}_g^{GI}$ are defined by
\begin{align}
\mO_g^{GI} & =\frac{i}{\Lambda^4}\bar{L}_L\cdot\phi e_R \left(\frac{\alpha_s}{4\pi}G_{\mu\nu}^a G^{a\mu\nu}\right)+\text{h.c.}
\label{eq:moggi} \\
\tilde{\mO}_g^{GI} & =\frac{1}{\Lambda^4}\bar{L}_L\cdot\phi e_R \left(\frac{\alpha_s}{4\pi}G_{\mu\nu}^a\tilde{G}^{a\mu\nu}\right)+\text{h.c.}
\label{eq:moggi2}
\end{align}
Here, $\Lambda$ is the energy scale at which new physics is integrated out. $L_L=
\begin{pmatrix} \nu_{eL} \\ e_L\end{pmatrix}$ is the left-handed lepton (electron) doublet while $e_R$
is the right-handed lepton (electron) singlet. $\phi$ is the SM Higgs doublet with the vacuum expectation value (vev) $\langle\phi\rangle=\frac{1}{\sqrt{2}}\begin{pmatrix} 0 \\ v\end{pmatrix}$ and $v=246\GeV$.
$G_{\mu\nu}^a=\partial_\mu G_\nu^a-\partial_\nu G_\mu^a+g_s f^{abc}G_\mu^b G_\nu^c$ denotes the
gluon field strength tensor, with $G_\mu^a$ being the gluon field, $f^{abc}$ the $SU(3)$ structure
constant, and $g_s$ the strong coupling constant with $\alpha_s=\frac{g_s^2}{4\pi}$. $\tilde{G}^{a\mu\nu}\equiv\frac{1}{2}\epsilon^{\mu\nu\rho\sigma}
G^{a}_{\rho\sigma}$ is the dual gluon field strength, with the convention $\epsilon^{0123}=+1$.
After electroweak symmetry breaking (EWSB), the $U(1)_{em}$-invariant CP-odd electron-gluonic
interactions can be written as
\begin{align}
\mL_{eg}^{CPV}\supset C_g\mO_g+\tilde{C}_g\tilde{\mO}_g
\label{eq:CO}
\end{align}
Here $\mO_g$ and $\tilde{\mO}_g$ are defined as
\begin{align}
\mO_g & =\frac{v}{\sqrt{2}\Lambda^4}\bar{e}i\gamma^5 e\left(\frac{\alpha_s}{4\pi}G_{\mu\nu}^a G^{a\mu\nu}\right) \\
\tilde{\mO}_g & =\frac{v}{\sqrt{2}\Lambda^4}\bar{e}e\left(\frac{\alpha_s}{4\pi}G_{\mu\nu}^a\tilde{G}^{a\mu\nu}\right)
\end{align}
Note through this definition the same set of Wilson coefficients
$C_g$ and $\tilde{C}_g$ appear in Eq.~\eqref{eq:OGI} and Eq.~\eqref{eq:CO}.
The introduction of the one-loop factor $\frac{\alpha_s}{4\pi}$ in the definition of the $\mO_{g}^{GI},
\tilde{\mO}_{g}^{GI},\mO_g,\tilde{\mO}_g$ has the advantage of simplifying the one-loop running behavior of
the operator coefficients. Moreover, if we imagine these operators are generated at one-loop
level with no helicity suppression, then we would naturally expect $C_g,\tilde{C}_g\sim\ord(1)$ (although
we will not confine ourselves to this possibility).
\begin{figure*}[ht]
\begin{centering}
\includegraphics[width=2.2in]{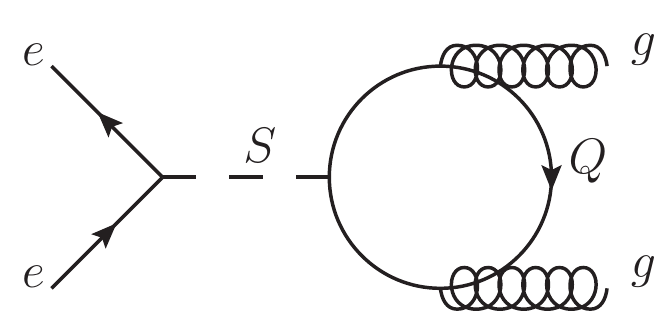} \\
\includegraphics[width=1.8in]{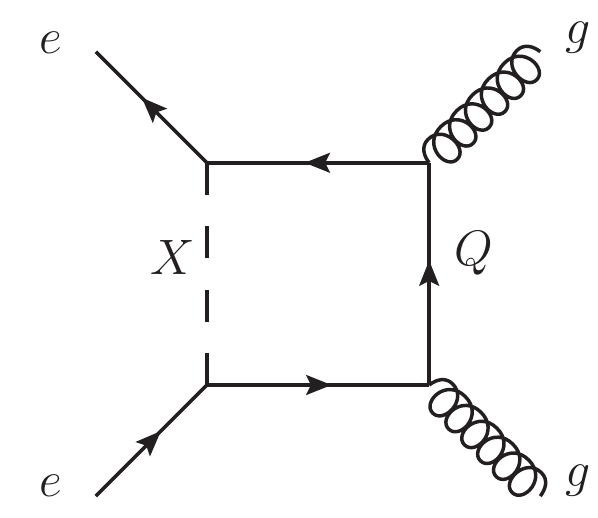}\quad\quad\quad\quad
\includegraphics[width=1.8in]{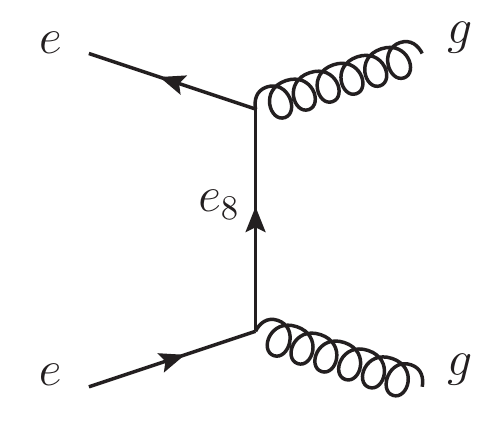}
\caption{\label{fig:eeggfd}Generation of effective electron-gluonic
operators in various perturbative UV realizations. Upper panel: Scalar-mediated toy model. Lower left panel: Leptoquark model. Lower right panel: $e_8$ model.}
\end{centering}
\end{figure*}

We note that the electron-gluonic interactions in Eq.~\eqref{eq:CO} can be generated
in several perturbative UV realizations. Three examples will be enumerated here, in which
we only intend to sketch the part of the model relevant for the electron gluon interactions,
with no attempt to provide a genuine viable UV completion. In the first example (the upper panel of
Fig.~\ref{fig:eeggfd}), a new color-singlet scalar $S$ is introduced, which has Yukawa couplings to both the SM electron
and a new quark $Q$ charged under the QCD color group, in a CP-violating manner, for instance
\begin{align}
\delta\mL_1=-Y_e\bar{e}i\gamma^5 eS-Y_Q\frac{m_Q}{\Lambda}\bar{Q}{Q}S
\end{align}
where for simplicity let us assume $\Lambda$ is the mass of $S$ and $m_Q$ is the mass of $Q$. Then we can estimate that
for $m_Q\gtrsim\Lambda$, $\mO_g$ is generated with coefficient $C_g\sim -\frac{\sqrt{2}\Lambda}{3v}Y_e Y_Q$. In the second
example (the lower left panel of Fig.~\ref{fig:eeggfd}), a scalar leptoquark $X$ is introduced,
which couples to the electron and a new quark $Q$ as follows
\begin{align}
\delta\mL_2=-\bar{e}(\lambda_{eQ}P_L+\lambda_{Qe}P_R)QX+\text{h.c.}
\end{align}
Here $P_{L/R}=\frac{1\mp\gamma^5}{2}$, and $\lambda_{eQ}=|\lambda_{eQ}|e^{i\theta_{eQ}},
\lambda_{Qe}=|\lambda_{Qe}|e^{i\theta_{Qe}}$ are complex coefficients. In this case $\mO_g$
is generated with coefficient $C_g\sim-\frac{\sqrt{2}\Lambda^2}{12vm_Q}|\lambda_{Qe}\lambda_{eQ}|\sin\Delta\theta$,
in which $\Lambda,m_Q$ denote the masses of $X$ and $Q$ respectively and $\Delta\theta\equiv\theta_{eQ}-\theta_{Qe}$.
In the third example, we consider a color octet electron~\cite{Sahin:2010dd,Akay:2010sw,Mandal:2012rx,Mandal:2016csb}, denoted $e_8$, which interacts with the electron and gluon
in the following CP-violating manner
\begin{align}
\delta\mL_3=\frac{g_s}{16\pi^2\Lambda}\bar{e}\sigma^{\mu\nu}(g_L P_L+g_R P_R)e_8^a G_{\mu\nu}^a
\end{align}
Here $\Lambda$ is the mass of $e_8$, and $g_L=|g_L|e^{i\theta_L},g_R=|g_R|e^{i\theta_R}$ are complex couplings.
When $e_8$ is integrated out, $\mO_g$ is generated with coefficient
$C_g\sim\frac{\sqrt{2}|g_L g_R|\sin\delta\theta\Lambda}{8\pi^2 v}$, in which $\delta\theta\equiv\theta_L-\theta_R$.
We note that in the second and the third example, the chirality flip does not occur on the electron line, therefore it
is natural that for these cases $C_g$ is not suppressed by the electron Yukawa. Moreover, $\tilde{O}_g$ can also
be generated (in the first example this would require more general complex couplings). We emphasize that although
we may have in mind certain perturbative UV realizations, we would like to adopt a model-independent approach
and thus be open-minded about the origins of new CP-odd electron-gluonic operators. In fact, for the perturbative
examples shown above, generically they lead to large direct contribution to electron EDM at one-loop order unless
we construct the models in some contrived or fine-tuned manner.

\section{Operator running and mixing}
\label{sec:orm}

$\mO_g$ and $\tilde{\mO}_g$ are expected to be generated by new physics at some
high scale (e.g. a few$\TeV$). However, EDM measurements are performed at a
much lower scale ($\lesssim 1\GeV$). In the effective field theory, operators
generally run and mix due to renormalization effects. Moreover, when going
below the threshold of some heavy particle, there could be matching corrections
when the heavy particle is integrated out. In this section we give an estimate
of the size of such effects, which lays the foundation for further analysis.

Let us start with the following CP-odd effective Lagrangian
\begin{align}
\mL_{eg}^{CPV}\supset C_g\mO_g+C_q\mO_q+\tilde{C}_g\tilde{\mO}_g+\tilde{C}_q\tilde{\mO}_q
\label{eq:CO2}
\end{align}
in which the operators $\mO_q$ and $\tilde{\mO}_q$ are defined by
\begin{align}
\mO_q & =\frac{v}{\sqrt{2}\Lambda^4}\bar{e}i\gamma^5 e(m_q\bar{q}q) \\
\tilde{\mO}_q & =\frac{v}{\sqrt{2}\Lambda^4}\bar{e}e(m_q\bar{q}i\gamma^5 q)
\end{align}
\begin{figure*}[ht]
\begin{centering}
\includegraphics[width=2.2in]{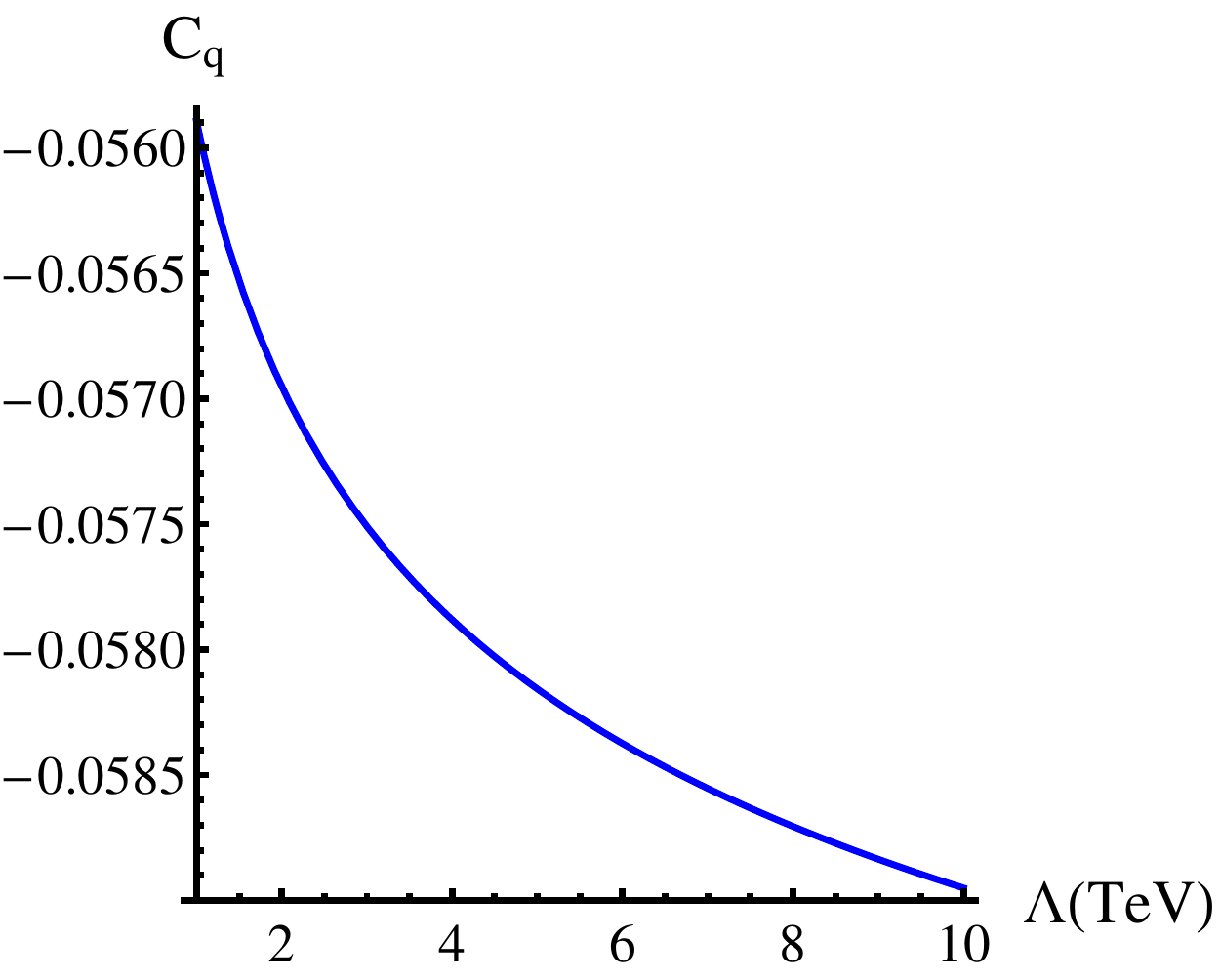}\quad\quad\quad\quad
\includegraphics[width=2.2in]{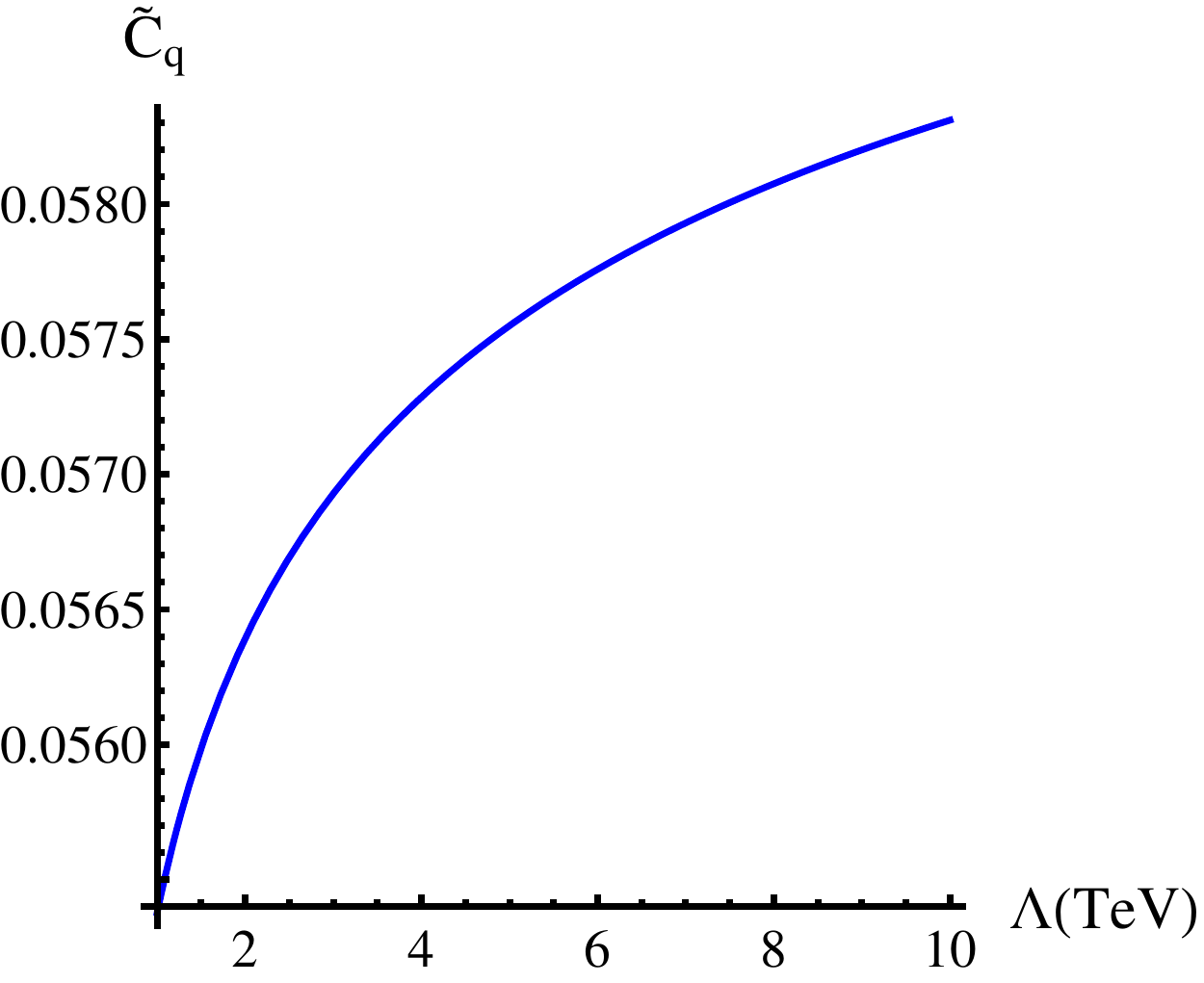}
\caption{\label{fig:cq} The induced Wilson coefficients $C_g$ (left) and
$\tilde{C}_g$ (right) for the operators ${\cal O}_g$ and
$\tilde{{\cal O}}_g$ of Eq. \eqref{eq:CO2} at the hadronic scale of $\mu=1~\textrm{GeV}$
versus the initial scale $\Lambda$, by numerically solving the set of equations
in Eq. \eqref{eq:beginrun} - Eq. \eqref{eq:endrun}, with the initial conditions $C_g=1,C_q =0$ (left) or
$\tilde{C}_g=1, \tilde{C}_q=0$ (right) at the scale $\Lambda$.}
\end{centering}
\end{figure*}
We include $\mO_q$ and $\tilde{\mO}_q$ since they appear in the running and
matching at (QCD) one-loop level. $m_q$ is the running mass of the quark $q$.
The Wilson coefficients $C_g,C_q,\tilde{C}_g,\tilde{C}_q$ are functions of the
renormalization scale $\mu$, with the leading order running behavior computed in the $\overline{\text{MS}}$
scheme to be
\begin{align}
\label{eq:beginrun}
\frac{d}{d\ln\mu^2}C_q & =\frac{\alpha_s^2}{\pi^2}C_g \\
\frac{d}{d\ln\mu^2}C_g & =0 \\
\frac{d}{d\ln\mu^2}\tilde{C}_q & = -\frac{\alpha_s^2}{\pi^2}\tilde{C}_g \\
\label{eq:endrun}
\frac{d}{d\ln\mu^2}\tilde{C}_g & =0
\end{align}
In the calculation, to preserve the anomaly equation of the singlet axial current and nonrenormalization
of the pseudoscalar quark operators we have introduced appropriate finite renormalization
as is done in ref.~\cite{Hill:2014yxa} (see also ~\cite{Larin:1993tq}). The results we obtained
for the running of $C_q$ and $\tilde{C}_q$ agree with ref.~\cite{Bhattacharya:2015rsa}.
\footnote{However, our running of $C_q$ is a factor of $4\pi$ smaller than
that given by the Appendix of ref.~\cite{Crivellin:2017rmk} in the context of
$\mu\rightarrow e$ conversion. Moreover, when
compared with ref.~\cite{Hill:2014yxa}, our running of $C_q$ agrees while the running of
$\tilde{C}_q$ obtained here is twice as large as that obtained by ref.~\cite{Hill:2014yxa}.}
Besides running, the Wilson coefficients
$C_g,\tilde{C}_g$ also receive matching corrections when the heavy quarks $q=c,b,t$ are integrated out
\begin{align}
C_g &\rightarrow C_g-\frac{C_q}{3}, \\
\tilde{C}_g &\rightarrow\tilde{C}_g+\frac{\tilde{C}_q}{2}
\end{align}
In this work, we take $2m_q$ to be the matching threshold.

In Fig.~\ref{fig:cq} we plot the induced Wilson coefficients $C_q$ (left) and $\tilde{C}_q$ (right) at $1\GeV$
scale assuming $C_g=1,C_q=0$ (or $\tilde{C}_g=1,\tilde{C}_q=0$) at scale $\Lambda$. In the calculation
the running of $\alpha_s$ is taken into account up to three-loop in QCD. For each flavor of light quark
$q=u,d,s$, the magnitude of the induced $C_q$ or $\tilde{C}_q$ is around 0.06, with a mild dependence
on $\Lambda$ for $\Lambda$ in a few$\TeV$ range.

\section{Observable EDM effects}
\label{sec:oee}
\begin{figure*}[ht]
\begin{centering}
\includegraphics[width=2.2in]{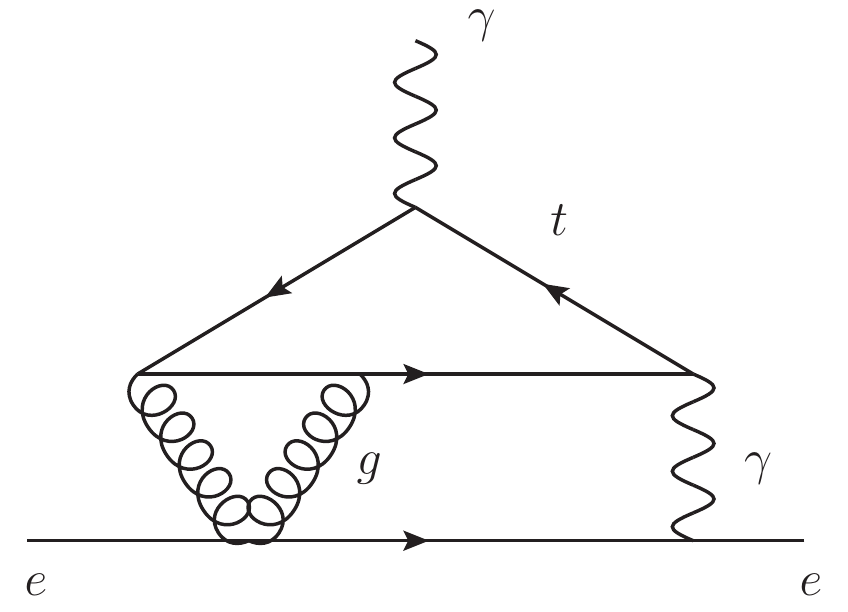}\quad\quad\quad\quad
\includegraphics[width=2.2in]{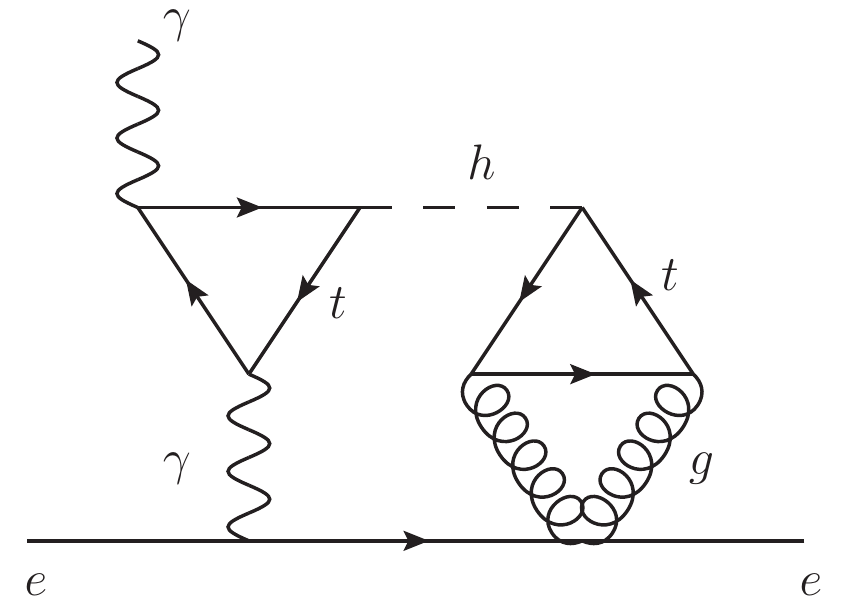}
\caption{\label{fig:loop} Representative Feynman diagrams of three (left)
or four (right) loop contribution to direct electron EDM from CP-odd
electron-gluonic operators. There is no two-loop diagrams contributing, because
the only diagram at two-loop level is to remove the internal photon line from the
left figure. However, this two-loop diagram is forbidden by Furry's theorem, because
after tracing the color indices, it is equivalent to the case in which three photons
contact to the same fermion loop.}
\end{centering}
\end{figure*}
Generally speaking, EDM measurements at molecular or atomic level are sensitive
to multiple sources of CP-violating effects\footnote{We refer the reader to
ref.~\cite{Khriplovich:1997ga,Pospelov:2005pr,Roberts:2010zz,Engel:2013lsa,
Jungmann:2013sga,Yamanaka:2014mda,Chupp:2017rkp,Safronova:2017xyt}
for reviews.}. For example, the ACME experiment
which uses ThO molecules is sensitive to both the electron EDM and CP-odd
electron-nucleon interactions. In this section we discuss how CP-violating
electron-gluonic operators may contribute to observable EDM effects.

\subsection{Direct electron EDM}

Furry's theorem prevents CP-odd electron-gluonic operators $\mO_g,\tilde{\mO}_g$
from contributing to the electron EDM at one-loop or two-loop order. Nevertheless,
at three-loop and four-loop order, we are able to draw diagrams, such as those shown in Fig.~\ref{fig:loop}.
The existence of such contributions can be easily understood. For example,
as shown in previous section, through renormalization group running $\mO_q$
can be induced from $\mO_g$. According to the analysis of operator mixing
patterns presented in ref.~\cite{Panico:2018hal} (see also~\cite{Cesarotti:2018huy}),
$\mO_q$ can generate the electron EDM operator at two-loop level. This effectively
leads to the three-loop diagram as shown in the left panel of Fig.~\ref{fig:loop}, which does not suffer
from helicity suppression. On the other hand, $\mO_g$ leads to an imaginary part
of the effective electron Yukawa coupling at two-loop level, which in turn generates
the electron EDM through a two-loop Barr-Zee diagram, leading to the four-loop
diagram shown in the right panel of Fig.~\ref{fig:loop}.

If $\mO_g$ is generated at scale $\Lambda$ with an $\ord(1)$ coefficient, then
its contribution to the electron EDM $d_e$ at three loop level can be estimated as
\begin{align}
|d_e|\sim\frac{\alpha_sm_t}{4\pi}\times\frac{\alpha_{em}}{4\pi}
\times\frac{ey_t}{16\pi^2}\times\frac{v^2}{2\Lambda^4}\times\frac{\alpha_s}{4\pi}
\approx\frac{3.6\times 10^{-29}}{(\Lambda/\textrm{TeV})^4}e\cdot\text{cm}
\label{eq:de1}
\end{align}
In the estimation expression above, the first three factors are loop factors, while
the last two factors come from vev insertion and operator definition. If this is the
only contribution to EDM measurements using ThO molecules then we are able to derive
a bound on $\Lambda$ from current constraint Eq.~\eqref{eq:acme} as
\begin{align}
\Lambda\gtrsim 1.3\TeV
\end{align}
For $\mO_g$ this turns out to be much weaker than the bound obtained from CP-odd electron-nucleon
interaction discussed below. Nevertheless, if $\tilde{\mO}_g$ is generated at scale $\Lambda$ with an $\ord(1)$ coefficient,
then we can also estimate the bound on $\Lambda$ as in the case of $\mO_g$ to be $\sim 1.3\TeV$.
This is comparable with or even more stringent than the bound from CP-odd electron-nucleon interaction discussed later.
We note that when an observable EDM receives contributions from both the direct electron EDM and CP-odd electron-nucleon
interactions, in principle we should consider these contributions simultaneously for setting a reliable bound
on the effective scale $\Lambda$, since there can be enhancement or cancellation effects. Nevertheless, because
the solution to three-loop running and matching equations is not yet available, in this paper we adopt a
simplistic approach by presenting separately the bounds obtained from considering only the direct electron EDM
contribution or only the CP-odd electron-nucleon interaction contribution. Therefore the bounds on effective scales
that we obtained by considering only one contribution should be interpreted as valid when the other contribution
is negligible, or be interpreted as an indication of the size of the corresponding contribution. According to the formulas
presented in the following subsections, for $\mO_g$, neglecting the direct electron EDM contribution is generally a safe
approximation. However, for $\tilde{\mO}_g$, the direct electron EDM contribution is not expected to be negligible,
therefore although we give bounds by considering two contributions separately, these bounds should be interpreted with
care.

The four-loop contribution to electron EDM (as shown in the right panel of Fig.~\ref{fig:loop})
can also be estimated. $\mO_g$ generated at $\Lambda$ with an $\ord(1)$ coefficient leads to an effective
imaginary part of the electron Yukawa
\begin{align}
|\text{Im}\kappa_e|\sim\frac{v}{\sqrt{2}\Lambda^4}\times\frac{\alpha_s}{4\pi}\times\frac{\alpha_s}{4\pi}\times
\frac{m_t^3}{16\pi^2}\times\frac{v}{\sqrt{2}m_e}
\end{align}
Here $\kappa_e$ denotes the electron Yukawa coupling relative to its SM value, thus we include a factor of
$\frac{v}{\sqrt{2}m_e}$ on the right hand side. $d_e$ receives a contribution from $|\text{Im}\kappa_e|$ via
a two-loop Barr-Zee diagram, and we estimate the contribution as~\cite{Altmannshofer:2015qra}
\begin{align}
|d_e|\sim 5.1\times|\text{Im}\kappa_e|\times 10^{-27}e\cdot\text{cm}
\approx\frac{8.7\times 10^{-31}}{(\Lambda/\textrm{TeV})^4}e\cdot\text{cm}
\end{align}
which is smaller by a factor of $40$ compared to the three-loop contribution estimated in Eq.~\eqref{eq:de1}.

Let us note that due to the gluon condensate $\langle 0|\frac{\alpha_s}{4\pi}G_{\mu\nu}^a G^{a\mu\nu}|0\rangle\sim\Lambda_{QCD}^4$
(and quark condensate $\langle 0|\bar{q}q|0\rangle$ if running is taken into account) the operator $\mO_g^{GI}$ will
also lead to corretion to $\text{Im}\kappa_e$ when we consider the condensate in Eq.~\eqref{eq:moggi} (the complex phase generated in
the electron mass term can be removed in the diagonalization of lepton mass matrix). However the correction is suppressed by
$\frac{\Lambda_{QCD}^4}{\Lambda^4}$ and thus too small to produce observable effects as long as $\Lambda\gtrsim 100\GeV$.

\subsection{CP-odd electron-Nucleon interaction}

The EDM measurements based on atoms and molecules may also be sensitive to CP-odd
electron-nucleon interactions. At hadron level, we parametrize the relevant operators as~\cite{Yamanaka:2014mda}
\begin{equation}
\mathcal{L}_{eN}\supset-\frac{G_F}{\sqrt{2}}\mathop{\sum}_{N=n,p}\left(C^{\textrm{SP}}_N\bar{N}N\bar{e}\textrm{i}\gamma^5e
+C^{\textrm{PS}}_N\bar{N}\textrm{i}\gamma^5N\bar{e}e\right)
\label{eq:hadron}
\end{equation}
In this work the tensor type electron-nucleon interaction induced from electron-gluonic operators is suppressed by
an electroweak loop factor and estimated to be negligible. When we perform a matching from quark level Eq.~\eqref{eq:CO2} to hadron level Eq.~\eqref{eq:hadron}
the coefficients $C^{\textrm{SP}}_N$ and $C^{\textrm{PS}}_N$ are obtained as follows
\begin{align}
C^{\textrm{SP}}_N &=-C_g\frac{v\left\langle\frac{\alpha_s}{4\pi}GG\right\rangle_N}{G_F\Lambda^4}
-\mathop{\sum}_{q=u,d,s}C_q\frac{v\left\langle m_q\bar{q}q\right\rangle_N}{G_F\Lambda^4},\\
C^{\textrm{PS}}_N &=-\tilde{C}_g\frac{2v\left\langle\frac{\alpha_s}{8\pi}G\tilde{G}\right\rangle_N}{G_F\Lambda^4}
-\mathop{\sum}_{q=u,d,s}\tilde{C}_q\frac{v\left\langle m_q\bar{q}i\gamma^5q\right\rangle_N}{G_F\Lambda^4}.
\end{align}
Here $\langle\mathcal{O}\rangle_N\equiv\langle N|\mathcal{O}|N\rangle$ denotes the nucleon matrix element of
the operator $\mO$ with respect to nucleon $N$. $GG$ and $G\tilde{G}$ are shorthand notations for $G^a_{\mu\nu}G^{a\mu\nu}$
and $G^a_{\mu\nu}\tilde{G}^{a\mu\nu}$, respectively. Note that the matching is expected to be performed at
the hadron scale $\sim 1\GeV$, and we integrate out heavy quarks $c,b,t$. Following the method in ref.~\cite{Hill:2014yxa,Cheng:2012qr},
we compute the relevant nucleon matrix elements as
\begin{align}
\left\langle\frac{\alpha_s}{4\pi}GG\right\rangle_p=\left\langle\frac{\alpha_s}{4\pi}GG\right\rangle_n=-183~\textrm{MeV}
\end{align}
\begin{align}
\left\langle\frac{\alpha_s}{8\pi}G\tilde{G}\right\rangle_p=-403~\textrm{MeV},\quad \left\langle\frac{\alpha_s}{8\pi}G\tilde{G}\right\rangle_n=31~\textrm{MeV}
\label{eq:nmem}
\end{align}
and the quark scalar and pseudoscalar matrix elements
\begin{align}
&\left\langle m_u\bar{u}u\right\rangle_p=15.5~\textrm{MeV},\quad
\left\langle m_u\bar{u}u\right\rangle_n=13.5~\textrm{MeV},\nonumber \\
&\left\langle m_u\bar{u}i\gamma^5u\right\rangle_p=383~\textrm{MeV},\quad
\left\langle m_u\bar{u}i\gamma^5u\right\rangle_n=-374~\textrm{MeV}, \nonumber \\
&\left\langle m_d\bar{d}d\right\rangle_p=29.4~\textrm{MeV},\quad
\left\langle m_d\bar{d}d\right\rangle_n=33.4~\textrm{MeV},\nonumber \\
&\left\langle m_d\bar{d}i\gamma^5d\right\rangle_p=-808~\textrm{MeV},\quad
\left\langle m_d\bar{d}i\gamma^5d\right\rangle_n=816~\textrm{MeV}, \nonumber \\
&\left\langle m_s\bar{s}s\right\rangle_p=40.2~\textrm{MeV},\quad
\left\langle m_s\bar{s}s\right\rangle_n=40.2~\textrm{MeV},\nonumber \\
&\left\langle m_s\bar{s}i\gamma^5s\right\rangle_p=-487~\textrm{MeV},\quad
\left\langle m_s\bar{s}i\gamma^5s\right\rangle_n=-54~\textrm{MeV}
\end{align}
Details about the computation are given in Appendix~\ref{sec:app}.
Here we note that the matrix elements of $\frac{\alpha_s}{8\pi}G\tilde{G}$ exhibit
significant isospin violation~\cite{Gross:1979ur}. Moreover, although the uncertainties
of most of the matrix elements listed above can be neglected in our analysis, the $1\sigma$ uncertainties
associated with the matrix elements of $\frac{\alpha_s}{8\pi}G\tilde{G}$ are estimated to be at least
$36\MeV$ for both the proton and the neutron, with a weak correlation. Especially for neutron this uncertainty is so
large that we conclude $\left\langle\frac{\alpha_s}{8\pi}G\tilde{G}\right\rangle_n$ is in fact
compatible with zero. This large uncertainty significantly weakens the bound on
$\tilde{\mO}_g$ operator.

\subsection{Observable EDM}

In this work we consider seven types of materials, including both paramagnetic
and diagmatic ones, to probe direct electron EDM and also various CP-violating electron-nucleon
interactions in a complementary manner. The seven types of materials are: ThO molecule, HfF$^+$ ion,
and atoms $^{205}\textrm{Tl}$,$^{199}\textrm{Hg}$,$^{129}\textrm{Xe}$,$^{211}\textrm{Rn}$,$^{225}\textrm{Ra}$.
Their observable EDMs are related to the electron EDM and CP-odd electron-nucleon interaction coefficients
as follows. For polar molecules~\cite{Chupp:2014gka,Cairncross:2017fip,Skripnikov:2016tho,Skripnikov:2017hff}:
\begin{align}
d^{\text{eff}}_{\text{ThO}} &=d_e-1.3\times 10^{-20}C_N^{\text{SP}}(e\cdot\text{cm}), \\
d^{\text{eff}}_{\text{HfF}} &=d_e-7.9\times 10^{-21}C_N^{\text{SP}}(e\cdot\text{cm}).
\end{align}
And for heavy atoms~\cite{Yamanaka:2014mda}\footnote{In recent years, some new results for the diamagnetic atoms appeared, see \cite{Yanase:2018qqq,Fleig:2018bsf,Sahoo:2018ile,Yamanaka:2017mef,Fleig:2018etq,Yoshinaga:2014eaa}. The new results are compatible with old ones
in order of magnitude, but some detailed behavior change much. As an example, the new estimations on the coefficients of $C^{\textrm{PS}}_p$ for $^{199}\textrm{Hg}$ and $^{129}\textrm{Xe}$ become ignorable. Such differences do not modify on our main conclusions.}:
\begin{align}
&d_{^{205}\textrm{Tl}}=-582d_e+\left(-7.0\times10^{-18}C^{\textrm{SP}}_N+1.8\times10^{-23}C^{\textrm{PS}}_p\right)~e\cdot\textrm{cm},\\
&d_{^{199}\textrm{Hg}}=-7.9\times 10^{-3}d_e
+\left(-5.1C^{\textrm{SP}}_N +0.61\left(0.09C^{\textrm{PS}}_p+0.91C^{\textrm{PS}}_n\right)\right)\times10^{-22}~e\cdot\textrm{cm},\\
&d_{^{129}\textrm{Xe}}=-0.98\times 10^{-3}d_e+\left(-6.2C^{\textrm{SP}}_N+1.6\left(0.24C^{\textrm{PS}}_p+0.76C^{\textrm{PS}}_n\right)\right) \times10^{-23}~e\cdot\textrm{cm},\\
&d_{^{211}\textrm{Rn}}=10.7\times 10^{-3}d_e+\left(7.0C^{\textrm{SP}}_N-0.71\left(0.02C^{\textrm{PS}}_p+0.98C^{\textrm{PS}}_n\right)\right) \times10^{-22}~e\cdot\textrm{cm},\\
&d_{^{225}\textrm{Ra}}=4.3\times 10^{-2}d_e+\left(2.9C^{\textrm{SP}}_N-0.64\left(0.25C^{\textrm{PS}}_p+0.75C^{\textrm{PS}}_n\right)\right) \times10^{-21}~e\cdot\textrm{cm}.
\end{align}
Note for ThO and HfF$^+$, since the experimental limit is usually given as a bound on the electron EDM,
we normalize the equation such that the coefficient in front of $d_e$ is unity. For simplicity, we neglect
isospin violating effects in $\text{SP}$ type electron-nucelon interactions, i.e., taking $C_N^{\text{SP}}
=C_n^{\text{SP}}=C_p^{\text{SP}}$.

\section{Current constraints and future prospects}
\label{sec:ccfp}
\subsection{Running and matching Effects}
\begin{figure*}[ht]
\begin{centering}
\includegraphics[width=1.8in]{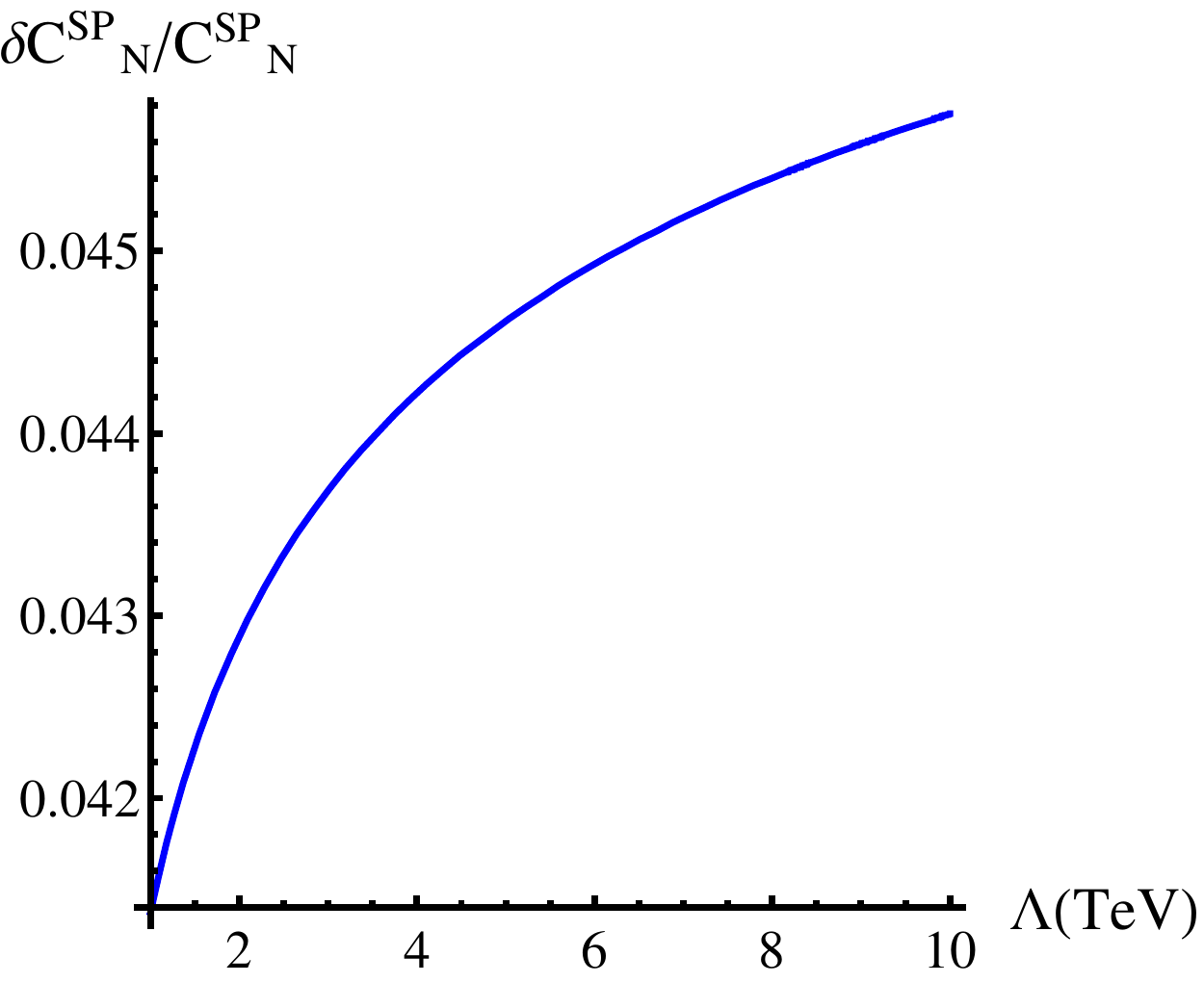}
\includegraphics[width=1.9in]{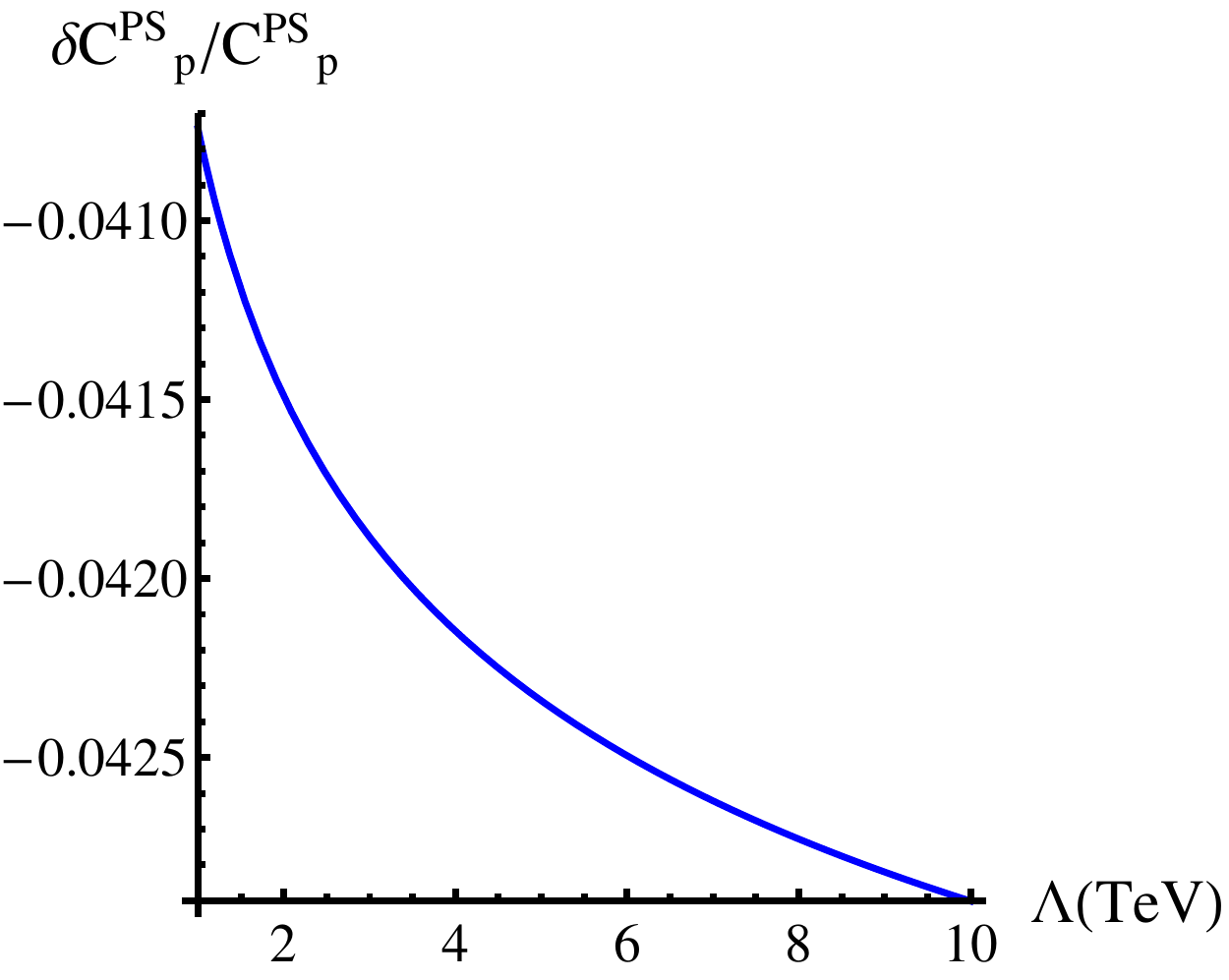}
\includegraphics[width=1.8in]{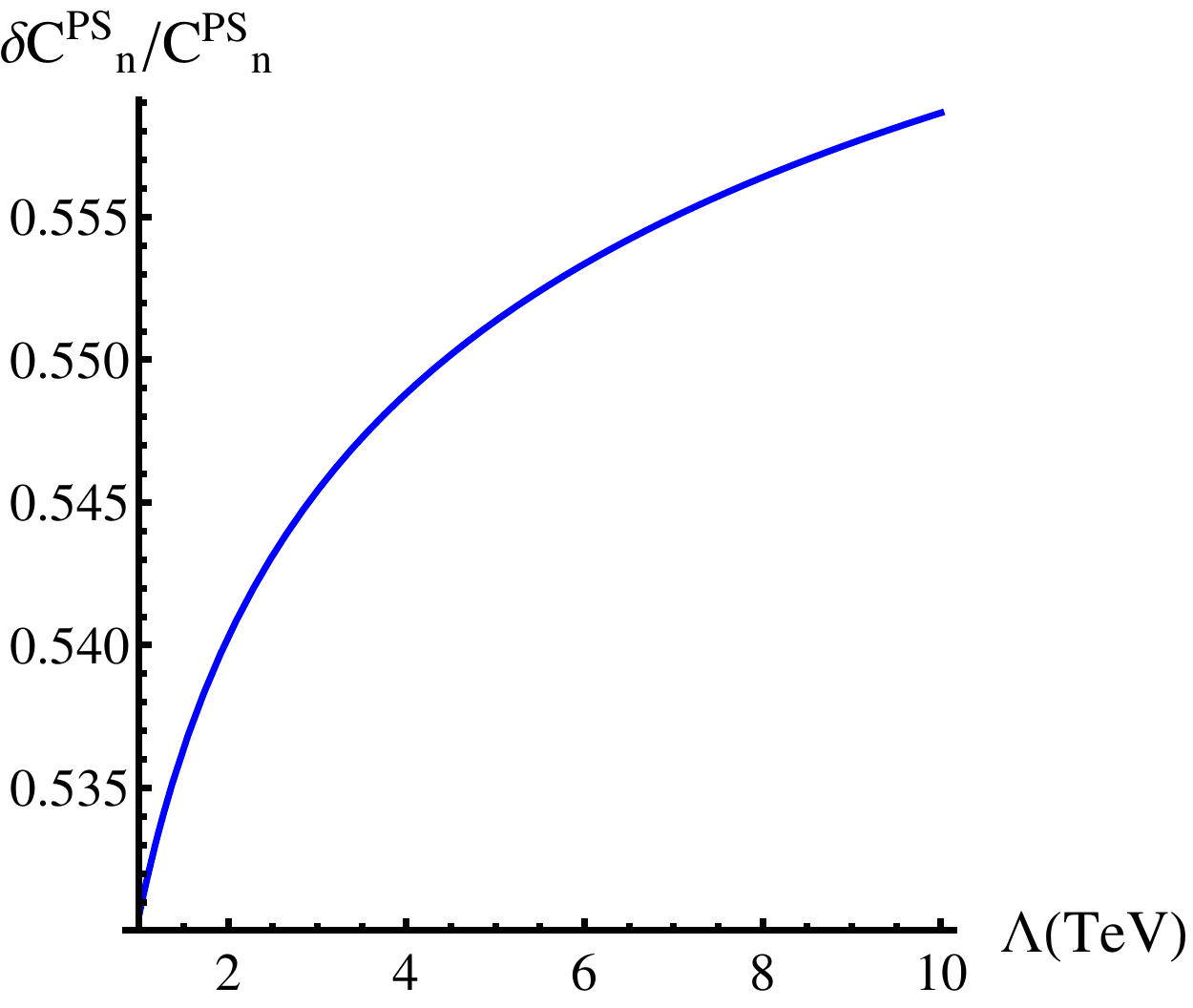}
\caption{\label{fig:delta} The relative corrections of the coefficients $C^{SP}_N$ and $C^{PS}_{p,n}$
of Eq.~\eqref{eq:hadron} due to QCD running and matching effects from
the initial scale $\Lambda$ to the hadronic scale $\mu=1~\textrm{GeV}$.
Left: assuming only a nonvanishing $C_g$ at $\Lambda$, with $N=p,n$;
middle and right: assuming only a nonvanishing $\tilde{C}_g$ at $\Lambda$.}
\end{centering}
\end{figure*}
Before deriving constraints on operator coefficients from various experiments, it is instructive
to obtain an estimate of the correction due to QCD running and matching effects by examining
hadron-level quantities which are closer to experimental observables than quark-level quantities
shown in Fig.~\ref{fig:cq}. In this process the nucleon matrix elements derived in Appendix~\ref{sec:app}
play a crucial role. In Fig.~\ref{fig:delta} we present the relative corrections due to QCD running and matching effects.
Assuming only $C_g$ or $\tilde{C}_g$ is generated at a high scale $\Lambda$, we may compute the coefficients
$C_N^{\text{SP}}$ or $C_p^{\text{PS}},C_n^{\text{PS}}$ at low scale $\mu\sim 1\GeV$ by taking into account or
neglecting the QCD running and matching effects discussed in Section~\ref{sec:orm}. The resulting differences
are denoted $\delta C_N^{\text{SP}},\delta C_p^{\text{PS}},\delta C_n^{\text{PS}}$ respectively, and in Fig.~\ref{fig:delta}
the relative corrections are obtained by dividing the corresponding coefficients without considering running and matching effects.

The important message coming from Fig.~\ref{fig:delta} is that all the CP-odd electron-proton interactions are quite insensitive
to the running and matching effects of the corresponding electron-gluonic operator (the correction is about $4\%$), while for CP-odd electron-neutron interactions, the SP type coefficient is insensitive but the PS type coefficient is sensitive. This can be traced to
the fact that the central value of the neutron matrix element $\langle\frac{\alpha_s}{8\pi}G\tilde{G}\rangle_n\approx31~\textrm{MeV}$ is much smaller than that
of proton $\langle\frac{\alpha_s}{8\pi}G\tilde{G}\rangle_p\approx403~\textrm{MeV}$, see Eq.~\eqref{eq:nmem}. Here we note that
since relative large uncertainty is associated with the neutron matrix element, the right panel of Fig.~\ref{fig:delta} also suffers
from large uncertainty. Nevertheless, we may expect the QCD running and matching effects to be significant when we want to constrain
$\tilde{C}_g$ using materials whose sensitivities are dominated by neutron rather than proton. This turns out to be the case for
$^{199}\textrm{Hg}$ and $^{211}\textrm{Rn}$. Unfortunately even after taking into account the running and matching effects, the uncertainties due to nucleon matrix elements are still too large to allow us to obtain meaningful bounds on $\tilde{C}_g$ from
these two materials.

\subsection{Contributions in benchmark scenarios}

In this section we fix $\Lambda=1\TeV$, and consider two benchmark scenarios: $C_g=1$ or $\tilde{C}_g=1$ at scale $\Lambda$.
We set coefficients of other CP-violating operators at $\Lambda$ to zero. We then consider the contribution to various observable
EDMs in these two scenarios. For simplicity, the contribution to direct electron EDM is not included. This is a good approximation
for constraining $C_g$ but we should keep in mind for constraining $\tilde{C}_g$ this effect is important for certain paramagnetic materials.
\begin{table*}[t!]
\begin{centering}
\begin{tabular}{|c|c|c|c|}
\hline
    & $C_N^{\text{SP}}$ & $C_p^{\text{PS}}$ & $C_n^{\text{PS}}$ \\
\hline
Tree-level & $3.86\times 10^{-6}$ & $1.70\times 10^{-5}$ & $-1.31\times 10^{-6}$ \\
\hline
One-loop running + matching & $4.02\times 10^{-6}$ & $1.63\times 10^{-5}$ & $-2.00\times 10^{-6}$ \\
\hline
\end{tabular}
\caption{The values of $C^{SP}_N$, $C^{PS}_p$, and $C^{PS}_n$ at hadronic
scale $\mu=1~\textrm{GeV}$ with and without taking into account the QCD running
and matching effects, with $N=p,n$. Here we take $C_g=1$ for the SP case and
$\tilde{C}_g=1$ for the PS case.}
\label{tab:bchhadron}
\end{centering}
\end{table*}

In Table~\ref{tab:bchhadron} we present the contribution to $C_N^{\text{SP}},C_p^{\text{PS}},C_n^{\text{PS}}$
(at low scale $\mu\sim1\GeV$) in the two benchmark scenarios, without or with running and matching effects.
Combined with the observable EDM formulae given in the previous section this facilitates a quick estimation
of typical contribution to observable EDMs from new CP-violation sources around$\TeV$ scale. In Table~\ref{tab:bcoedm}
we present the estimated contribution to various observable EDMs in the two benchmark scenarios.

\begin{table*}[t!]
\begin{centering}
\begin{tabular}{|c|c|c|c|}
\hline
Observable$(e\cdot\textrm{cm})$ & $C_g=1$ & $\tilde{C}_g=1$ & $\tilde{C}_g=1$ (R)\\
\hline
$d^{\text{eff}}_{\text{ThO}}$ & $-5.2\times 10^{-26}$ & - & - \\
\hline
$d^{\text{eff}}_{\text{HfF}}$ & $-3.2\times 10^{-26}$ & - & - \\
\hline
$d_{^{205}\textrm{Tl}}$ & $-2.8\times10^{-23}$ & $(3.1\pm0.3)\times10^{-28}$ & $(2.9\pm0.3)\times10^{-28}$ \\
\hline
$d_{^{199}\textrm{Hg}}$ & $-2.0\times10^{-27}$ & $(2.1\pm8.4)\times10^{-29}$ & $(-2.2\pm8.4)\times10^{-29}$ \\
\hline
$d_{^{129}\textrm{Xe}}$ & $-2.5\times10^{-28}$ & $(4.9\pm1.9)\times10^{-29}$ & $(3.8\pm1.9)\times10^{-29}$ \\
\hline
$d_{^{211}\textrm{Rn}}$ & $2.8\times10^{-29}$ & $(0.7\pm1.1)\times10^{-28}$ & $(1.2\pm1.1)\times10^{-28}$ \\
\hline
$d_{^{225}\textrm{Ra}}$ & $1.2\times10^{-26}$ & $(-2.1\pm0.7)\times10^{-27}$ & $(-1.6\pm0.7)\times10^{-27}$ \\
\hline
\end{tabular}
\caption{Contributions to observable EDMs in benchmark scenarios described in the text. Note: contribution to direct electron EDM is not included. The results in the column marked with (R) are computed with QCD running and matching effects which are not taken into
account in the remaining columns. For $\tilde{C}_g=1$ scenario, the indicated uncertainties are conservative estimates
based on the estimation of uncertainties of nucleon matrix elements given in Appendix~\ref{sec:app}.}
\label{tab:bcoedm}
\end{centering}
\end{table*}

\begin{table*}[t!]
\begin{centering}
\begin{tabular}{|c|c|c|}
\hline
$|d_{\text{eN}}/d_{\text{eEDM}}|$ & $C_g=1$ & $\tilde{C}_g=1$ \\
\hline
$\text{ThO}$ & $1.5\times 10^3$ & $0$ \\
\hline
$\text{HfF}^+$ & $8.8\times 10^2$ & $0$ \\
\hline
$^{205}\textrm{Tl}$ & $1.3\times 10^3$ & $1.4\times 10^{-2}$ \\
\hline
$^{199}\textrm{Hg}$ & $7.2\times 10^3$ & $7.6\times 10^1$$^*$ \\
\hline
$^{129}\textrm{Xe}$ & $7.1\times 10^3$ & $1.1\times 10^3$$^*$ \\
\hline
$^{211}\textrm{Rn}$ & $7.3\times 10^3$ & $3.0\times 10^2$$^*$ \\
\hline
$^{225}\textrm{Ra}$ & $7.5\times 10^3$ & $1.1\times 10^3$$^*$ \\
\hline
\end{tabular}
\caption{Estimate of the ratio of the contribution from CP-odd electron-nucleon interaction to the contribution of direct electron EDM
in benchmark scenarios described in the text. The numbers should all be viewed as rough estimates. For entries marked with an asterisk,
the ratio might be significantly overestimated, considering the large uncertainties shown in Table~\ref{tab:bcoedm}.}
\label{tab:eNratio}
\end{centering}
\end{table*}

For the observable EDM measurement from any given material, it is also instructive to compare the contribution
from CP-odd electron-nucleon interaction $d_{\text{eN}}$ and the contribution from the direct electron EDM $d_{\text{eEDM}}$.
For the two benchmark scenarios, we present the ratio $|d_{\text{eN}}/d_{\text{eEDM}}|$ in Table~\ref{tab:eNratio}, which
can be calculated from the results in Table~\ref{tab:bcoedm} and the direct electron EDM estimate in Eq.~\eqref{eq:de1}
(the various enhancement or suppression factors for electron EDM should be taken into account if relevant). The results
shown in Table~\ref{tab:eNratio} should be viewed as rough order-of-magnitude estiamtes, but it still delivers useful information.
Let us note that in forming the ratio, the factor $\frac{C_g}{\Lambda^4}$ or $\frac{\tilde{C}_g}{\Lambda^4}$ will cancel (when the running
effect is not significant), therefore the value of the factor $\frac{C_g}{\Lambda^4}$ or $\frac{\tilde{C}_g}{\Lambda^4}$ is not important
for estimating the ratio. Then from Table~\ref{tab:eNratio} we may conclude that for $C_g\neq 0$ scenario, the contribution
from CP-odd electron-nucleon interaction always dominate, and it is safe to neglect the direct electron EDM contribution.
However, for $\tilde{C}_g\neq 0$ scenario, it is hard to identify any one material in which the contribution from CP-odd
electron-nucleon interaction will certainly dominate, considering the large uncertainties shown in Table~\ref{tab:bcoedm}.

\subsection{Constraints from current and expected experiments}
\begin{table*}[t!]
\begin{centering}
\begin{tabular}{|c|c|c|c|c|}
\hline
Observable$(e\cdot\textrm{cm})$ & Current Limit & Confidence Level & Future Limit & Ref. \\
\hline
$|d^{\text{eff}}_{\text{ThO}}|$ & $1.1\times 10^{-29}$ & $90\%$ & $10^{-31}$ & ~\cite{Andreev:2018ayy,Vutha:2009ux} \\
\hline
$|d^{\text{eff}}_{\text{HfF}}|$ & $1.3\times 10^{-28}$ & $90\%$ & $10^{-30}$ & ~\cite{Cairncross:2017fip} \\
\hline
$|d_{^{205}\textrm{Tl}}|$ & $9.5\times 10^{-25}$ & $90\%$ & - & ~\cite{Regan:2002ta} \\
\hline
$|d_{^{199}\textrm{Hg}}|$ & $7.4\times 10^{-30}$ & $95\%$ & - & ~\cite{Graner:2016ses} \\
\hline
$|d_{^{129}\textrm{Xe}}|$ & $5.5\times 10^{-27}$ & $90\%$ & $3\times 10^{-29}$ & ~\cite{Chupp:2014gka,Engel:2013lsa}\\
\hline
$|d_{^{211}\textrm{Rn}}|$ & - & - & $2\times 10^{-28}$ & ~\cite{Engel:2013lsa} \\
\hline
$|d_{^{225}\textrm{Ra}}|$ & $1.4\times 10^{-23}$ & $95\%$ & $10^{-28}$ & ~\cite{Bishof:2016uqx} \\
\hline
\end{tabular}
\caption{Current and future experimental upper limits on various observable EDMs.
The confidence level refers to current limits.}
\label{tab:expedm}
\end{centering}
\end{table*}

In Table~\ref{tab:expedm} we present current and future (expected) experimental upper limits on various observable EDMs.
A comparison between Table~\ref{tab:bcoedm} and Table~\ref{tab:expedm} gives a feeling about the constraining power
on $C_g$ and $\tilde{C}_g$ for fixed $\Lambda$, or equivalently, on $\Lambda$ for fixed $C_g$ and $\tilde{C}_g$,
bearing in mind that the new physics contribution roughly scale as $\Lambda^{-4}$. In Table~\ref{tab:cfb}
the current and future experimental upper limits on various observable EDMs are translated into lower bounds
on the effective scale $\Lambda$ in the definition Eq.~\eqref{eq:moggi} and Eq.~\eqref{eq:moggi2}.
Specifically, the $\Lambda$ column refers to the current lower bound assuming $C_g=1$, the $\tilde{\Lambda}$
column refers to the current lower bound assuming $\tilde{C}_g=1$, the $\Lambda^*$ column refers to
the expected future lower bound assuming $C_g=1$, and the $\tilde{\Lambda}^*$ column refers to the expected future
lower bound assuming $\tilde{C}_g=1$. We consider one parameter at a time and thus set coefficients of other
new physics operators at the effective scale to zero. Note the bound is obtained by neglecting the contribution
to direct electron EDM via three-loop running or matching. The confidence levels are the same as those shown in
Table~\ref{tab:expedm} for current limits while for future projection since the sensitivities are all rough estimates
we do not distinguish between $90\%$ and $95\%$ at the moment. Entries marked with a dash mean no meaningful bound
can be obtained due to large uncertainties or lack of data or information of future experiments. We have neglected
nucleon matrix element uncertainties in obtaining these bounds, which could lead to large uncertainties in
the bounds on $\tilde{\Lambda}$ and $\tilde{\Lambda}^*$ for $^{129}\textrm{Xe}$
and $^{225}\textrm{Ra}$. We put the corresponding results in parentheses to indicate their being afflicted by
large nucleon matrix element uncertainties.

\begin{table*}[t!]
\begin{centering}
\begin{tabular}{|c|c|c|c|c|}
\hline
Observable & $\Lambda(\text{TeV})$ & $\tilde{\Lambda}(\text{TeV})$ & $\Lambda^*(\text{TeV})$ & $\tilde{\Lambda}^*(\text{TeV})$ \\
\hline
$|d^{\text{eff}}_{\text{ThO}}|$ & 8.3 & - & 27 & - \\
\hline
$|d^{\text{eff}}_{\text{HfF}}|$ & 4.0 & - & 13 & - \\
\hline
$|d_{^{205}\textrm{Tl}}|$ & 2.3 & 0.13 & - & - \\
\hline
$|d_{^{199}\textrm{Hg}}|$ & 4.1 & - & - & - \\
\hline
$|d_{^{129}\textrm{Xe}}|$ & 0.46 & (0.29) & 1.7 & (1.1) \\
\hline
$|d_{^{211}\textrm{Rn}}|$ & - & - & 1.9 & - \\
\hline
$|d_{^{225}\textrm{Ra}}|$ & 0.17 & (0.1) & 3.3 & (2.0) \\
\hline
\end{tabular}
\caption{Lower bound on the effective scale $\Lambda$ in the definition Eq.~\eqref{eq:moggi} and Eq.~\eqref{eq:moggi2}.
See text for detailed explanation.}
\label{tab:cfb}
\end{centering}
\end{table*}

According to Table~\ref{tab:cfb}, currently the most stringent bound on the effective scale of $(\bar{e}i\gamma^5 e)G_{\mu\nu}^a G^{a\mu\nu}$ comes from the ThO experiment, which gives a lower bound of about $8\TeV$. In the future this bound could be improved
to $27\TeV$. These bounds are impressive in that the electron-gluonic operator is dimension-eight in the context of SMEFT.
For $(\bar{e}e)G_{\mu\nu}^a \tilde{G}^{a\mu\nu}$, currently a relatively reliable bound comes from $^{205}\textrm{Tl}$, which
set the effective scale to be larger than about $0.13\TeV$. Bounds from other materials suffer from large nucleon matrix element uncertainties. As discussed in Section~\ref{sec:oee} considering the contribution to direct electron EDM generated by three-loop running and matching could set a lower bound on the effective scale for $(\bar{e}e)G_{\mu\nu}^a \tilde{G}^{a\mu\nu}$ as $\sim 1.3\TeV$ from current ThO experiment. This could be further improved to $\sim 4\TeV$ in the future. However, bounds obtained in this way are currently only order-of-magnitude estimates, which suffer from a large uncertainty unless the solution to the associated running and matching equations is obtained.

\section{Discussion and conclusions}
\label{sec:dnc}

In this paper we have examined current and future bounds on the CP-odd electron-gluonic operators $(\bar{e}i\gamma^5 e)G_{\mu\nu}^a G^{a\mu\nu}$ and $(\bar{e}e)G_{\mu\nu}^a \tilde{G}^{a\mu\nu}$, or in the gauge invariant form shown in Eq.~\eqref{eq:moggi}
and Eq.~\eqref{eq:moggi2}. They could arise from integrating out CP-violating new physics at a high scale $\Lambda$, say, a few$\TeV$.
In the SMEFT framework they are represented as dimension-eight operators and thus their effects are suppressed by $\Lambda^{-4}$.
Nevertheless we found the current ThO experiment can already put an impressive lower bound on the effective scale of the operator
$(\bar{e}i\gamma^5 e)G_{\mu\nu}^a G^{a\mu\nu}$ at around $8\TeV$. This is obtained by assuming the parametrization in Eq.~\eqref{eq:moggi}
and thus implicitly we are assuming the new physics contribution does not suffer from helicity suppression (otherwise the bound would be
significantly weakened). We have also shown explicitly that QCD running effects can bring only a mild correction to the bound on the coefficient
of $(\bar{e}\textrm{i}\gamma^5e)G_{\mu\nu}G^{\mu\nu}$ operator which was extracted from ACME II ThO measurement, thus we can obtain a similar bound through only a tree level analysis. Future ThO measurements are expected to push the bound to $27\TeV$. One interesting aspect about the operator $(\bar{e}i\gamma^5 e)G_{\mu\nu}^a G^{a\mu\nu}$ is that it contributes to observable EDM mainly through CP-odd electron nucleon interaction
rather than direct electron EDM (which only arises from three-loop level). For the operator $(\bar{e}e)G_{\mu\nu}^a \tilde{G}^{a\mu\nu}$, if we consider only the contribution from CP-odd electron-nucleon interactions, there is no constraint from merely ACME experiment, because the averaged spin for both kind of nucleons vanish in Th- and O-nuclei. Even after considering other materials,
we still found the current bound on the effective scale is weak ($\sim 0.13\TeV$ from $^{205}\textrm{Tl}$). Its contribution to direct electron EDM via three-loop running and matching might give a more stringent bound ($\sim 1.3\TeV$), but this only serves as an order-of-magnitude estimate.

Drell-Yan process at hadron colliders could also probe lepton-gluonic operators, as studied in ref.~\cite{Potter:2012yv,Hayreter:2013vna,Cai:2018cog}.
A detailed collider study for the CP-odd electron-gluonic
operators is beyond the scope of the present work. Nevertheless, from the results of ref.~\cite{Cai:2018cog}
for lepton-flavor-violating lepton-gluonic operators we could estimate current LHC sensitivity to the effective
scale is at best around $1\TeV$. For the operator $(\bar{e}i\gamma^5 e)G_{\mu\nu}^a G^{a\mu\nu}$, the collider sensitivity is
certainly not comparable with ThO experiments. However for the operator $(\bar{e}e)G_{\mu\nu}^a \tilde{G}^{a\mu\nu}$,
the collider sensitivity might be comparable and a detailed simulation (and also detailed study of the uncertainties in predicting the observable EDMs) is needed to determine whether collider experiments could deliver a more stringent constraint. It should be noted that
in any case, the collider probe is not sensitive to the CP-nature of the interaction, unlike EDM observables. We therefore expect
future EDM experiments, hopefully involving new materials and techniques, could play an indispensable role in discovering or
constraining new sources of CP-violation.

\acknowledgments

We thank Yi-Lei Tang for helpful discussions about the $m_t$ dependence of the three-loop contribution to direct electron EDM.
We also thank Jordy de Veris and Nodoka Yamanaka for helpful discussions about the nucleon matrix elements and the EDM at atomic level.
This work was supported in part by the MoST of Taiwan under the grants no.:105-2112-M-007-028-MY3 and 107-2112-M-007-029-MY3.

\begin{appendix}
\section{Nucleon Matrix Elements}\label{sec:app}
In this appendix we outline the derivation of relevant nucleon matrix elements which are important
for relating the quark-level Lagrangian to the hadron-level Lagrangian. We first consider the matrix
element $\left\langle\frac{\alpha_s}{4\pi}GG\right\rangle_N$, which appears in the following sum rule
after performing the heavy quark expansion~\cite{Ji:1994av,Cheng:2012qr,Hill:2014yxa}
\begin{align}
m_N=\left(1+\frac{2\alpha_s}{\pi}\right)\mathop{\sum}_{q=u,d,s}\left\langle m_q\bar{q}q\right\rangle_N-\frac{9}{2}\left\langle\frac{\alpha_s}{4\pi}GG\right\rangle_N.
\label{eqn:sumrule}
\end{align}
We note that $\pi N$ and strange $\sigma$ terms (for both $N=p,n$) have been obtained by lattice calculation as~\cite{Yang:2015uis}
\footnote{There are also some results from other lattice groups or chiral perturbation calculation, as summarized in \cite{Yamanaka:2018uud}. The results vary in the region $\sigma_{\pi N}\sim(30-60)~\textrm{MeV}$, and thus the result we quoted in this paper is close to the averaged value. Its variance modify the final gluon matrix elements at percent level, which means our final results are not sensitive to this variance.}
\begin{align}
\sigma_{\pi N}&\equiv\frac{m_u+m_d}{2}\left\langle\bar{u}u+\bar{d}d\right\rangle_N=(45.9\pm7.9)~\textrm{MeV},\\
\sigma_s&\equiv\left\langle m_s\bar{s}s\right\rangle_N=(40.2\pm12.2)~\textrm{MeV}.
\end{align}
On the other hand, another combination of quark matrix elements are obtained in ref.~\cite{Crivellin:2013ipa}
with the aid of $SU(2)$ chiral perturbation theory
\begin{align}
\sigma_{-,p}&\equiv(m_d-m_u)\langle\bar{u}u-\bar{d}d\rangle_p=(2\pm1)~\textrm{MeV},\\
\sigma_{-,n}&\equiv(m_d-m_u)\langle\bar{u}u-\bar{d}d\rangle_n=(-2\pm1)~\textrm{MeV},
\end{align}
If we define $\Sigma_N\equiv\sum_{q=u,d,s}\left\langle m_q\bar{q}q\right\rangle_N$, then from the above results
for $\sigma_{\pi N},\sigma_s,\sigma_{-,p},\sigma_{-,n}$ we can obtain
\begin{align}
\Sigma_p&=\sigma_{\pi N}+\sigma_s-\frac{\sigma_{-,p}}{2}=(85.1\pm14.6)~\textrm{MeV},\\
\Sigma_n&=\sigma_{\pi N}+\sigma_s-\frac{\sigma_{-,n}}{2}=(87.1\pm14.6)~\textrm{MeV}.
\end{align}
These results allow us to obtain $\left\langle\frac{\alpha_s}{4\pi}GG\right\rangle_N$ from Eq.~\eqref{eqn:sumrule}
\begin{align}
\left\langle\frac{\alpha_s}{4\pi}GG\right\rangle_p=(-183.2\pm4.3)~\textrm{MeV},
\quad\left\langle\frac{\alpha_s}{4\pi}GG\right\rangle_n=(-182.9\pm4.3)~\textrm{MeV}.
\end{align}
The difference between proton and neutron is very small and thus neglected when this nucleon matrix
element is involved in the calculation.

Let us then turn to the estimation of $\left\langle\frac{\alpha_s}{8\pi}G\tilde{G}\right\rangle_N$.
We introduce the axial-vector currents
\begin{align}
A_0^{\mu}&\equiv\frac{1}{3}\left(\bar{u}\gamma^{\mu}\gamma^5u+\bar{d}\gamma^{\mu}\gamma^5d+\bar{s}\gamma^{\mu}\gamma^5s\right),\\
A_3^{\mu}&\equiv\frac{1}{2}\left(\bar{u}\gamma^{\mu}\gamma^5u-\bar{d}\gamma^{\mu}\gamma^5d\right),\\
A_8^{\mu}&\equiv\frac{1}{2\sqrt{3}}\left(\bar{u}\gamma^{\mu}\gamma^5u+\bar{d}\gamma^{\mu}\gamma^5d-2\bar{s}\gamma^{\mu}\gamma^5s\right).
\end{align}
The associated form factors at zero momentum transfer are then defined by
\begin{align}
\mathop{\lim}_{q\rightarrow0}\left\langle N|A^{\mu}_a|N\right\rangle=F_a^N\bar{u}^N\gamma^{\mu}\gamma^5u^N.
\end{align}
in which $u^N$ denotes the nucleon spinor. From the divergence of the above equation we obtain
\begin{align}
F_0^Nm_N\left(\bar{u}^N\textrm{i}\gamma^5u^N\right)&=\frac{1}{3}\sum_{q=u,d,s}\left\langle m_q\bar{q}\textrm{i}\gamma^5q\right\rangle_N-\left\langle\frac{\alpha_s}{8\pi}G\tilde{G}\right\rangle_N,\\
F_3^Nm_N\left(\bar{u}^N\textrm{i}\gamma^5u^N\right)&=\frac{1}{2}\left(\left\langle m_u\bar{u}\textrm{i}\gamma^5u\right\rangle_N-\left\langle m_d\bar{d}\textrm{i}\gamma^5d\right\rangle_N\right),\\
F_8^Nm_N\left(\bar{u}^N\textrm{i}\gamma^5u^N\right)&=\frac{1}{2\sqrt{3}}\left(\left\langle m_u\bar{u}\textrm{i}\gamma^5u\right\rangle_N+\left\langle m_d\bar{d}\textrm{i}\gamma^5d\right\rangle_N-2\left\langle m_s\bar{s}\textrm{i}\gamma^5s\right\rangle_N\right).
\end{align}
In the following we will drop the factor $\left(\bar{u}^N\textrm{i}\gamma^5u^N\right)$. To facilitate the comparison with
experiments, we introduce $\Delta q^N,N=p,n$, defined as
\begin{align}
\Delta q^N\equiv\langle N|\bar{q}\gamma_\mu\gamma^5 q|N\rangle s^\mu
\end{align}
where $s^\mu$ represents the nucleon spin 4-vector. It is proven in ref.~\cite{Dienes:2013xya} that
$\Delta q^N$ satisfies
\begin{align}
m_N\Delta q^N=\left\langle m_q\bar{q}\textrm{i}\gamma^5q\right\rangle_N-\left\langle\frac{\alpha_s}{8\pi}G\tilde{G}\right\rangle_N
\end{align}
This allows us to obtain relations between $F_a^N$ and $\Delta q^N$
\begin{align}
2F^p_3 &=\Delta u^p-\Delta d^p=-2F_3^n,\\
2\sqrt{3}F^p_8 &=\Delta u^p+\Delta d^p-2\Delta s^p=2\sqrt{3}F^n_8,
\end{align}
From the review on axions in ref.~\cite{Patrignani:2016xqp} we take the following experimental values
\begin{align}
2F^p_3&=1.269\pm0.003,\\
2\sqrt{3}F^p_8&=0.586\pm0.031,\\
\Delta s^p&=-0.09\pm0.02=\Delta s^n.
\end{align}
Then we are able to obtain
\begin{align}
\Delta u^p=-\Delta d^n=0.84\pm0.03,\quad\quad \Delta d^p=-\Delta u^n=-0.43\pm0.03.
\end{align}
To obtain the gluon matrix element, it is traditional to adopt the large $N_c$ chiral limit, implying the
constraint (with uncertainty at $\ord(N_c^{-1})$ understood)~\cite{Cheng:2012qr,Hill:2014yxa,Dienes:2013xya}
\begin{align}
\mathop{\sum}_{q=u,d,s}\langle\bar{q}\textrm{i}\gamma^5q\rangle_N=0
\end{align}
Then we are able to obtain
\begin{align}
\left\langle\frac{\alpha_s}{8\pi}G\tilde{G}\right\rangle_N&=-m_N\frac{\sum_q(\Delta q^N/m_q)}{\sum_q(1/m_q)}\nonumber\\
&=-m_N\left(\Delta s^N+\frac{F_3^N(1-m_u/m_d)}{1+m_u/m_d+m_u/m_s}+\frac{\sqrt{3}F_8^N}{1+(m_s/m_u+m_s/m_d)^{-1}}\right).
\end{align}
Numerically we have
\begin{align}
\tilde{G}_p\equiv\left\langle\frac{\alpha_s}{8\pi}G\tilde{G}\right\rangle_p=(-403\pm36)~\textrm{MeV},\quad\quad
\tilde{G}_n\equiv\left\langle\frac{\alpha_s}{8\pi}G\tilde{G}\right\rangle_n=(31\pm36)~\textrm{MeV},
\end{align}
with a correlation $R=-0.15$. Recent lattice data reduced the mass ratio uncertainty to $m_u/m_d=(0.46\pm0.05)$ \cite{Tanabashi:2018oca,Aoki:2016frl}, thus the uncertainties from $m_u/m_d$, $F_8^N$, and $\Delta s^N$ are of the same order $\sim\mathcal{O}(10~\textrm{MeV})$. The dominant part of uncertainties still comes from $m_u/m_d$ which leads to $|\delta\tilde{G}_{n,p}|\sim27~\textrm{MeV}$. However, it should be kept in mind that the uncertainties of nucleon
matrix elements given here are intended as conservative estimates since several sources of uncertainties are not taken into
account. Quark matrix elements and their uncertainty estimates can also be obtained from the relations presented above. We display
the corresponding results in Table~\ref{tab:qm1} and Table~\ref{tab:qm2}. Note that due to correlations, in these tables
the uncertainty of the sum of matrix elements can be smaller than the uncertainty of the matrix element of an individual quark flavor.
\begin{table*}[t!]
\begin{centering}
\begin{tabular}{|c|c|c|c|c|}
\hline
Quark $q$&$u$&$d$&$s$&sum\\
\hline
$\left\langle m_q\bar{q}q\right\rangle_p$ (MeV) & $15.5\pm2.7$ & $29.4\pm5.5$ & $40.2\pm12.2$ & $85.1\pm14.6$ \\
\hline
$\left\langle m_q\bar{q}q\right\rangle_n$ (MeV) & $13.5\pm2.7$ & $33.4\pm5.5$ & $40.2\pm12.2$ & $87.1\pm14.6$ \\
\hline
\end{tabular}
\caption{Nucleon matrix elements of scalar quark operators $m_q\bar{q}q$.}
\label{tab:qm1}
\end{centering}
\end{table*}
\begin{table*}[t!]
\begin{centering}
\begin{tabular}{|c|c|c|c|c|}
\hline
Quark $q$&$u$&$d$&$s$&sum\\
\hline
$\left\langle m_q\bar{q}\textrm{i}\gamma^5q\right\rangle_p$ (MeV)&$383\pm39$&$-808\pm27$&$-487\pm31$&$-912\pm31$\\
\hline
$\left\langle m_q\bar{q}\textrm{i}\gamma^5q\right\rangle_n$ (MeV)&$-374\pm39$&$816\pm27$&$-54\pm31$&$388\pm31$\\
\hline
\end{tabular}
\caption{Nucleon matrix elements of pseudoscalar quark operators $m_q\bar{q}i\gamma^5 q$.}
\label{tab:qm2}
\end{centering}
\end{table*}

\end{appendix}


\bibliography{eegg_v7}
\bibliographystyle{JHEP}

\end{document}